\documentclass[aps,prl,twocolumn,showpacs,groupaddress, footinbib]{revtex4-1}
\usepackage{amsmath}
\usepackage{amsfonts}
\usepackage{multirow}
\usepackage{float}
\usepackage{graphicx}
\usepackage{subfigure}
\usepackage{xcolor}
\newcommand{\obrt}{\frac{1}{\sqrt{2}}}
\newcommand{\lp}{\left(}
\newcommand{\rp}{\right)}
\newcommand{\diff}{\mathrm{d}}

\newcommand{\fact}[1]{#1!}

\newcommand{\ket}[1]{\vert #1 \rangle}

\newcommand{\ie}{\emph{i.e. }}
\newcommand{\viz}{\emph{viz.}}

\newcommand{\via}{\emph{via }}
\begin{document}
\title{Long-Range Repulsion Between Spatially Confined van der Waals Dimers}
\author{Mainak Sadhukhan}
\author{Alexandre Tkatchenko}
\email{alexandre.tkatchenko@uni.lu}
\affiliation{Physics and Materials Science Research Unit, University of Luxembourg, L-1511 Luxembourg}
\date{\today}
\begin{abstract}
It is an undisputed textbook fact that non-retarded van der Waals (vdW) interactions between isotropic dimers are attractive, regardless of the polarizability of the interacting systems or spatial dimensionality. The universality of vdW attraction is attributed to the dipolar coupling between fluctuating electron charge densities. Here we demonstrate that the long-range interaction between \textit{spatially confined} vdW dimers becomes repulsive when accounting for the full Coulomb interaction between charge fluctuations. Our analytic results are obtained by using the Coulomb potential as a perturbation over dipole-correlated states for two quantum harmonic oscillators embedded in spaces with reduced dimensionality, however the long-range repulsion is expected to be a general phenomenon for spatially-confined quantum systems. We suggest optical experiments to test our predictions, analyze their relevance in the context of intermolecular interactions in nanoscale environments, and rationalize the recent observation of anomalously strong screening of the lateral vdW interactions between aromatic hydrocarbons adsorbed on metal surfaces.
\end{abstract}
\pacs{Intermolecular potentials and forces, 34.20.Gj,
Harmonic oscillators, 03.65.Ge,
Adsorption on solid surface, 68.43.-h,
Perturbation theory, applied to atomic physics, 31.15.xp
}
\maketitle
Interactions induced by quantum-mechanical charge density fluctuations, 
such as van der Waals (vdW) and Casimir forces, are always present 
between objects with finite dimensions~\cite{parsegian2005van,Klimchitskaya2009,DobsonJPCM2012,Woods2016rmp}. 
Such interactions are important not only for many fundamental phenomena
throughout the fields of biology, chemistry, and physics, but also for the   design 
and performance of micro- and nano-structured devices. 
While Casimir forces can be both attractive or repulsive, depending on the nature of
quantum/thermal fluctuations and topology/geometry of the interacting 
systems~\cite{Zhao2009,Rodriguez2011,Rodriguez-Lopez2014,Jakubczyk2016}, it is
an undisputed common wisdom that non-retarded vdW interactions between two 
objects \textit{in vacuo} are inherently attractive~\cite{Langbein1974,stone1997theory,kaplan2006intermolecular}.
The universality of vdW attraction is attributed to the ubiquitous zero-point energy lowering 
induced by dipolar coupling between fluctuating electron charge densities~\cite{Langbein1974,stone1997theory}.

However, many biological, chemical, and physical phenomena of importance in
materials happen in spatially confined environments, as opposed to isotropic and homogeneous vacuum. 
The confinement can be artificially engineered by applying static or dynamic electromagnetic fields, 
or arise as a result of encapsulation of molecules in nanotubes, fullerenes, and/or by
adsorption on polarizable surfaces. Moreover, in biological systems, proteins are typically confined 
in an inhomogeneous environment. We remark that even when such confinement 
entails tiny modification of the electron density (having no apparent effect on the electrostatics), 
it can visibly affect the interactions stemming from density fluctuations due to their long-range 
inhomogeneous nature.

Here we demonstrate that the breaking of rotational and/or translational symmetry of 3D vacuum
results in repulsive long-range interactions for vdW dimers.
The repulsive interaction stems from the full Coulomb coupling between
charge density fluctuations, and is a universal signature of constrained electric-field lines in
1D, 2D, or quasi-3D spaces. In fact, reported cases of long-range repulsion between physisorbed
molecules abound in recent experimental 
literature~\cite{Wagner2010, Kleimann2014,Liu2014a, Stadtmuller2015, Thussing2016}.
The usual explanation attributes the repulsion to the charge transfer between the Fermi level 
of the metal surface and the molecular orbitals of the adsorbate~\cite{Simpson2012} or the dominance 
of Pauli repulsion over London-type dispersion interaction~\cite{Kroger2011, Tabor2011, Shajesh2012}.
These explanations, however, do not apply to large molecules physisorbed on metallic surfaces.
Our calculations suggest an alternative explanation for these and other experiments in nano-confined
systems. In contrast to previously known cases of van der Waals repulsions, which are either mediated by another molecule (three-body Axilrod-Teller-Muto interaction~\cite{Axilrod1943}) or a dielectric medium~\cite{Tabor2011}, the present effect has a different distance dependence and is general under arbitrary confinement.

\begin{figure}[h!]
 \includegraphics[width=0.42 \textwidth]{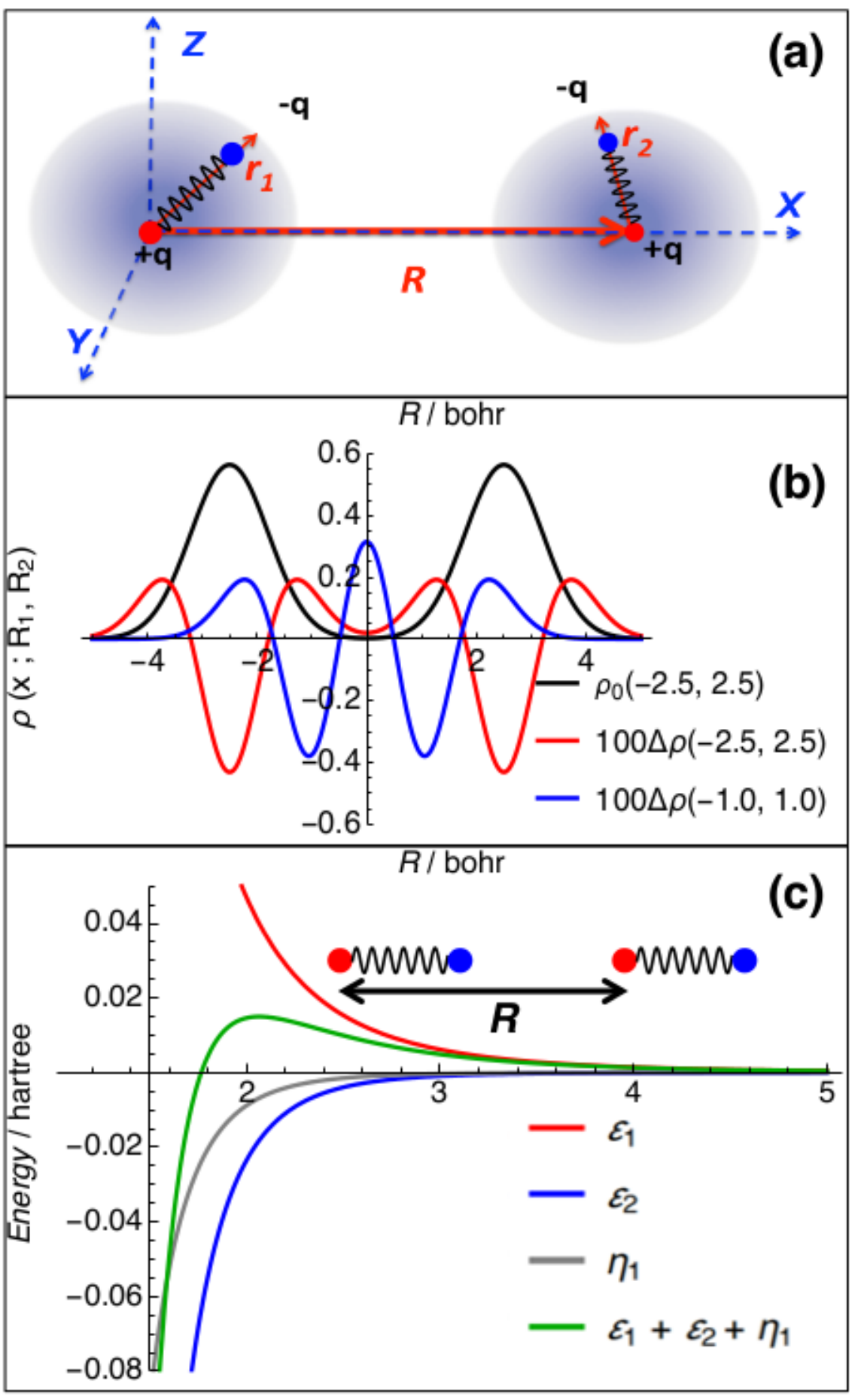}
\caption{(a) Drude oscillators embedded in the global reference frame (blue). (b) Probability density of uncoupled oscillators $\rho_{0}$ placed at $\pm 2.5$ \emph{bohr} (black), the difference density ($\Delta \rho = \rho_{dip} - \rho_{0}$) of dipole-correlated oscillators (red) and the $\Delta \rho$ for the pair of oscillators (blue) placed at $\pm 1$ \emph{bohr}. (c) Leading order interaction terms of beyond-dipole perturbation expansion in 1D for $q=m=\omega = 1$.}
  \label{fig:dipolestate}
\end{figure}

We start our analysis by investigating a pair of coupled isotropic 3D Drude 
oscillators~\cite{Bade1957,Bade1957a,Bade1958} (charge-separated, overall-neutral 
quantum harmonic oscillators) in reduced spatial dimensions (see Fig.~\ref{fig:dipolestate}(a) and discussion below). The Drude oscillators model the instantaneous, quantum-mechanical electronic fluctuations (not the permanent deformations of the electron density), therefore being a model for electron correlation \emph{via} the adiabatic connection fluctuation-dissipation theorem \cite{langreth1975}. Consequently, the analysis of this work corresponds to dynamic electron correlation effects in confined environments. 
Despite being bosonic, Drude oscillators~\cite{Bade1957} provide a reliable and robust model
of van der Waals interactions between valence electron 
densities~\cite{Wang2001,Sommerfeld2005,Tkatchenko2009,Jones2009,Tkatchenko2012,Gobre2013,Voora2013,Jones2013,Lopes2013,Liu2014,Ambrosetti2014,Ferri2015a,Odbadrakh2015,Sokhan2015, sadhukhan16}.
For example, the harmonic oscillator model, within the so-called many-body dispersion (MBD) framework, 
has been applied to accurately model vdW interactions in 
molecules, molecular crystals, solids with and without defects, surfaces, and 
nanostructured materials~\cite{Gobre2013,TkatchenkoAFM2015,Reillychemsci2015,Maurer_jcp2015,wang2015prl,wang2013prl}. 
Here, we go beyond the MBD model by developing a perturbative analysis to study the effects of the full Coulomb 
interaction on the physically confined electric-field lines of the dipole-coupled Drude oscillators. 

The developments in this article are based on the perturbation theory of Coulomb-coupled Drude oscillators developed by Jones et al. in Ref.\cite{Jones2013}. We also develop and apply an alternative perturbation expansion, which takes coupled dipolar oscillators as a starting point. 

Individual Drude oscillators follow quantum mechanics of harmonic oscillators of frequency $\omega$ and mass $\mu$. The full Coulomb potential between a pair of Drude oscillators, each connected to charges $\pm q$ (using vacuum permittivity $\epsilon_{0}= 1/4\pi$) is  
\begin{equation}\label{totcoul}
  V = q^2
      \left\{\frac{1}{|\mathbf R|}
           + \frac1{|\mathbf R-\mathbf r_1+\mathbf r_2|}
           - \frac1{|\mathbf R-\mathbf r_1|}
           - \frac1{|\mathbf R+\mathbf r_2|}
      \right\}       ,
\end{equation}
where $\mathbf R$, $\mathbf r_1$ and $\mathbf r_2$ refer to the interoscillator separation vector and individual oscillator coordinates (positive charges are origins of indivial oscillators' coordinates), respectively (see Fig.~\ref{fig:dipolestate}(a)). The zero-distance limit for $\mathbf r_1$ and $\mathbf r_2$, a presumably valid approximation at large $\mathbf{R}$, results in a widely used dipole-approximated potential
\begin{equation}
  V_\text{dip} = \frac{q^2}{R^5}
        \Big\{ R^2\,\mathbf r_1\!\cdot\!\mathbf r_2
               - 3\,(\mathbf r_1\!\cdot\!\mathbf R)\,(\mathbf r_2\!\cdot\!\mathbf R)
        \Big\}   ,          
\end{equation}
which allows an exact solution of the coupled oscillator problem, and leads to an interoscillator attraction regardless of the oscillator parameters and dimensionality of space. A general 3D oscillator state $\ket{\mathbf{n}}$ (total quanta $n$) is the product of three independent 1D oscillators 
\begin{equation}
 \langle  \zeta \ket{\mathbf{n}} = \frac{1}{\sqrt{2^{n_{\zeta}} \fact{n_{\zeta}}}}\left( \frac{\mu_{\zeta} \omega_{\zeta}}{\pi \hbar}\right)^{1/4}e^{-\left[\frac{\mu_{\zeta}\omega_{\zeta} {{\zeta}}^2}{2\hbar}\right]} H_{n_{\zeta}}\left( \sqrt{\frac{\mu_{\zeta} \omega_{\zeta}}{\hbar}}{\zeta} \right)
\end{equation} 
where $\zeta \in \{x,y,z\}$, $n =\sum_{\zeta} n_{\zeta}$ and $H_{n_{\zeta}}$ is a Hermite polynomial of order $n_{\zeta}$. Therefore, {any energy integral corresponding to Eq.\eqref{totcoul} for a 3D Drude oscillator} can be expressed as the product of three independent 1D integrals due to the identity 
\begin{equation}
  \frac{1}{r} = \frac{2}{\sqrt{\pi}}\int_{0}^{\infty} e^{-s^{2}r^{2}} \diff s .
\end{equation}
Now we will consider the limiting case of confinement as complete quenching of oscillator motion in one or more direction which is equivalent to negligible polarizations in those directions. A mathematical equivalent of fully-quenched motion of one of the oscillator components (say $Z$) can be obtained \via $n_{z} = 0$ and $\mu_z \to \infty$. Consequently, the 3D model transforms to an effective 2D model following the identity $\delta(x) = \lim_{a \to \infty} \lp\frac{a}{\sqrt{\pi}}\rp e^{-a^{2}x^{2}}$. Similarly, we can obtain an effective 1D model by {completely} confining two dimensions of the original 3D model. Resulting effective quasi-1D/2D potentials vary with the inverse of interoscillator distance (will be called ``Coulomb'' potential henceforth) but do not satisfy the Laplace equation for the oscillator charges, unlike the original 3D model. Physically, the restricted motions confine the electric-field lines in a restricted region of space, which produces repulsion between the oscillators, as we will show below.

{We start by analyzing the quantum mechanics of a full-Coulomb-coupled 1D oscillator dimer.} Formally, a Taylor expansion of {the interoscillator (with individual coordinates $x_1$ and $x_2$) 1D ``Coulomb'' potential is
\begin{equation}\label{1Dpotential}
  V_{int}=  \sum_{m = 2}^{\infty} V_{m} =\frac{q^2}{R} \sum_{m = 2}^{\infty}\sum_{k = 1}^{m-1} \binom{m}{k} \frac{x_{1}^{m-k} (-x_{2})^{k}}{R^m}.
\end{equation}
Full-Coulomb-coupled oscillator states can be obtained by perturbing independent oscillator states with $V_{int}$ starting with dipole potential given by $m = 2$ term in Eq.\eqref{1Dpotential}} (\emph{full-Coulomb perturbation}) or, perturbing the analytic dipole-coupled oscillator states \cite{Odbadrakh2015} by $V_{int}$ starting from $m = 3$ of Eq.\eqref{1Dpotential} (\emph{beyond-dipole perturbation expansion}). While these {two} approaches are formally equivalent, in practice the beyond-dipole expansion converges faster and allows novel insights into Coulomb-coupled oscillators (see \cite{Note1}).

	A system of two similar (mass $\mu$ and frequency $\omega$), dipole-coupled, 1D Drude oscillators is equivalent to two independent oscillators \cite{Bade1957,Bade1957a,Bade1958} in collective coordinates
	\begin{eqnarray}
	 a_1 = \frac{x_1 + x_2}{\sqrt{2}} \label{collcoord1}\\
	 a_2 = \frac{x_1 - x_2}{\sqrt{2}} \label{collcoord2}
	 \end{eqnarray}
	 with frequencies $\omega_1 = \omega / f_{-}$ and $\omega_2 = \omega / f_{+}$, respectively where
\begin{equation}\label{f1f2definition}
f_{\pm} = \lp1\pm\frac{2q^2}{\mu\omega^2 R^3}\rp^{-1/2}.
\end{equation} 
Figure~\ref{fig:dipolestate}(b) shows the anisotropic charge density that is created around free oscillator states due to dipole coupling, leading to emerging dipole moments. The consequent lowering of zero-point energy contains London attraction ($-C_6/R^6$) as the leading contribution but does not contain effects due to higher-order multipole moments~\cite{Odbadrakh2015}. {We now employ \emph{beyond-dipole perturbation expansion} on these dipole-coupled states as defined in the previous paragraph. Two leading order perturbation terms correspond to $m = 3$ and $4$ in Eq.\eqref{1Dpotential} and when expressed in collective coordinates $a_1$ and $a_2$ (Eqs.\eqref{collcoord1},\eqref{collcoord2}), they are }
\begin{equation}\label{v31d}
V_{3} = \frac{-3 q^2}{\sqrt{2} R^4} \lp a_{1}^{2} a_{2} - a_{2}^{3} \rp
\end{equation}
 and
\begin{equation}\label{v41d}
V_{4} = \frac{q^2}{2R^5} \lp 7a_2^4 - a_1^4 - 6 a_1^2 a_2^2 \rp ,
\end{equation}
 repectively. 
{In the ground state of dipole-coupled oscillators has even symmetry with respect to $a_1$ and $a_2$, the leading contribution to the first order beyond-dipole perturbation theory comes from leading even-order potential in Eq.\eqref{1Dpotential} \viz $V_{4}$  leading to first-order correction from forth-order potential (Eq.\eqref{v41d}) (superscript and subscript on energy components indicate order of perturbation and order of potential, respectively)}
\begin{equation}\label{e1_4_1d}
E^{(1)}_{4} = \frac{3q^2}{8R^5}\lp \frac{\hbar}{\mu \omega}\rp^2 \left[ 7 f_{+}^2 - f_{-}^2 - 2 f_{-}f_{+} \right]
\end{equation}  
{Similar symmetry consideration yields the leading second order perturbation correction from Eq.\eqref{v31d} (see S.I. for details) }
\begin{equation}\label{e2_3_1d}
   E^{(2)}_{3} = -\frac{9q^4}{16 \mu\omega^2 R^8}\lp \frac{\hbar}{\mu \omega}\rp^{2}\left[  3f_{+}^2 - f_{+}f_{-}\right]^2 .
\end{equation} 
	The power series expansion of Eq.~\eqref{e1_4_1d} in powers of $R$ (using Eq.~\eqref{f1f2definition}) yields
{
\begin{equation}
E^{(1)}_{4} = \varepsilon_1 + \varepsilon_2 + \ldots = 2 \frac{\alpha_2  \hbar\omega}{R^5} - 8  \frac{\alpha_2 \alpha_{1}   \hbar \omega}{R^8} + \dots
\end{equation}
}
The appearance of quadrupole polarizability $\alpha_2 =  \frac{3}{4}\lp \frac{q^2}{\mu \omega^2}\rp \lp \frac{\hbar}{\mu \omega}\rp$ in the {leading} repulsive term 
{
\begin{equation}\label{leadingrep}
 \varepsilon_1 = 2  \frac{\alpha_2  \hbar\omega}{R^5}
\end{equation}
}
and the following attractive term 
{
\begin{equation}
\varepsilon_2 = - 8 \frac{\alpha_2 \alpha_{1}  \hbar \omega}{R^8}
\end{equation}
}
indicates that the {$\varepsilon_1$} corresponds to the mean-field energy of an instantaneous quadrupole in the field generated by fluctuations in another electronic fragment. A similar approach yields 

\begin{equation}
\eta_1 = = -3 \frac{\alpha_{2}\alpha_{1}  \hbar\omega}{R^8}
\end{equation}

 as the leading term of Eq.\eqref{e2_3_1d}. Note that {$\eta_1$} appears to share similar physical origin with {$\varepsilon_2$ \ie both  $\eta_1$ and $\varepsilon_2$ come} from the interaction between the singly-excited and the ground states of the noninteracting oscillator pair. {Similar power law could have also come from the first-order perturbation correction due to $V_{7}$, which however vanishes identically.} As expected, the cumulative effect of {$\varepsilon_1$, $\varepsilon_2$ and $\eta_1$} remains repulsive (Fig.~\ref{fig:dipolestate}(c)) in the long range. The addition of the London dispersion $C_6 R^{-6}$ term does not alter the asymptotic repulsion. {Note, the source of this repulsion is $\varepsilon_1$, which is} proportional to the quadrupole polarizability (linear in $\hbar$) unlike the case of isotropic and homogeneous vacuum (see Fig.7 in Ref.~\cite{Jones2013}).

 It is important to note that $E^{(1)}_{4}$, which originates from the fourth 
derivative of the $V_{int}$, vanishes identically in isotropic and homogeneous vacuum unlike the aforementioned quasi-1D case. The analogous quantity in the full-3D case
\begin{equation}\label{fourthderivative}
 \sum_{\alpha, \beta, \gamma, \delta} \frac{\partial^4}{\partial r_{\alpha} \partial r_{\beta} \partial r_{\gamma} \partial r_{\delta}} V_{Coul}
\end{equation}
sums over four variables $\alpha$, $\beta$, $\gamma$ and $\delta$, each of which can have only 3 values \ie $x$, $y$ and $z$. As a result, all possible combinations of $\alpha, \beta, \gamma$ and $\delta$ in Eq.~\eqref{fourthderivative} 
contain at least one repeated index, yielding
\begin{equation}\label{fourthdervative1}
 \sum_{ \gamma, \delta}\frac{\partial^2}{\partial r_{\gamma}\partial r_{\delta}} \sum_{\alpha} \frac{\partial^2}{\partial r_{\alpha}\partial r_{\alpha}} V_{coul}.
\end{equation}
The second summation in Eq.~\eqref{fourthdervative1} vanishes, resulting in the dipole potential as the correct asymptotic limit in the case of homogeneous and isotropic vacuum.



The extension of our analysis to two dimensions is straightforward. Similar to Eq.\eqref{1Dpotential}, the Laplace expansion of the Coulomb potential, followed by the power series expansion of the Legendre polynomial yields the even parity terms of interoscillator interaction as
\begin{equation}
\begin{split}
&V^{(2p)} = \frac{4^{p}q^{2}}{R^{2p+1}}\sum_{t=0}^{p}\sum_{s=0}^{p-t}\binom{2p}{2t}\binom{\frac{2(p+t)-1}{2}}{2p}\binom{p-t}{s} \\
& \times \left[ (\Delta x)^{2(p-s)} (\Delta y)^{2s} - x_{2}^{2(p-s)} y_{2}^{2s} -x_{1}^{2(p-s)} y_{1}^{2s}\right]
\end{split}
\end{equation}
where $\Delta x = (x_1-x_2)$ and $\Delta y = (y_1-y_2)$ and $p \in \mathbb{Z}^{+}$ .
 Aligning $\mathbf R$ with the global $X$ direction { \ie along the line joining positive charges of two oscillators},
\begin{equation}\label{v4_2d}
\begin{split}
&V^{(4)} = V^{(4)}_{x} + \frac{3}{4}V^{(4)}_{y} - \\
&\frac{3q^2}{2R^{5}}\left[ 7a_{2}^2 b_{2}^{2}-4a_{1} b_{1}a_{2}b_{2} -a_{1}^2 b_{1}^{2}- a_{1}^2 b_{2}^{2}-a_{2}^2 b_{1}^{2} \right]
\end{split}
\end{equation}
is obtained as the fourth order term in collective coordinates $a_{1} = \obrt ( x_{1} + x_{2}), a_{2} = \obrt ( x_{1} - x_{2}), b_{1} = \obrt ( y_{1} + y_{2})$ and $b_{2} = \obrt ( y_{1} - y_{2})$. The $Y$-directional component $V^{(4)}_{y} = \frac{\tilde{q}^2}{R^{5}} \left[ 6y_{1}^{2}y_{2}^{2}-4\lp y_{1}^{3}y_{2} + y_{2}^{3}y_{1} \rp\right]$ is expressed in terms of reduced charge $\tilde{q} = q/\sqrt{2}$ to retain the usual form of dipole term. {The third term in Eq.~\eqref{v4_2d} is the coupling between $X$ and $Y$ components of the potential}. The final expression shows that the leading contribution to the long-range energy
\begin{equation}\label{e1_4_2d}
 E^{(1)}_{2D, 4} \approx \frac{3}{4}\varepsilon_{1,y} = \frac{3}{2} \frac{\alpha_{2}^{y}  \hbar \omega}{R^5}
\end{equation}
arises out of the cancellation between the leading orders of the repulsive first and the attractive third term of Eq.~\eqref{v4_2d}. The resulting long-range repulsive energy, similar to the 1D example, is the mean-field energy of an instantaneous quadrupole and varies as $R^{-5}$. Curiously, the predominant effect comes from the orthogonal component (to the {interoscillator-axis}, here global $X$ axis) of polarizability and may have non-trivial implications, for example for molecules confined between layered materials~\cite{nairnatcomm2016}. 

Next, we extend our analysis to the quasi-3D case, which is relevant for modeling lateral interactions between
molecules adsorbed on polarizable (metallic or semiconducting) surfaces.
For molecules physisorbed on surfaces, the interaction potential follows the Poisson equation. 
Aligning the global $X-Y$ plane with the 
surface (here we assume perfect reflection and discuss the general case later) and $X$-axis with $\mathbf R$, 
one realizes that the $Z$-component of the oscillator experiences a half-oscillator potential
\begin{equation}
\begin{split}
V(z) = \begin{cases} 
      \infty & \textrm{; } z \le 0  \\
      \frac{m \omega^2}{2} z^2& \textrm{; } z > 0 \\
   \end{cases} 
\end{split}
\end{equation}
The resulting oscillator state will be the product of usual $X$ and $Y$-directional 1D oscillators and the 
\begin{equation}
 \langle z \ket{\mathbf{n}} = N_{n_z} H_{2n_z+1}\left(\sqrt{\frac{m \omega}{\hbar}}z\right)e^{-\frac{m\omega z^2}{2 \hbar}} \Theta(z)
\end{equation}
with $N_{n_z} = \frac{1}{\sqrt{4^{n_z} \fact{(2n_z+1)} }}\lp \frac{m \omega}{\pi\hbar}\rp^{1/4}$ and $\Theta(z)$ is the step function accounting for the absence of oscillator wavefunction below the metal surface. The continuity of the wavefunction therefore demands a node on the surface forcing the solution to be the odd subset of full quantum oscillator solutions. 

\begin{figure}[h!]
\begin{center}
\includegraphics[width=0.4 \textwidth]{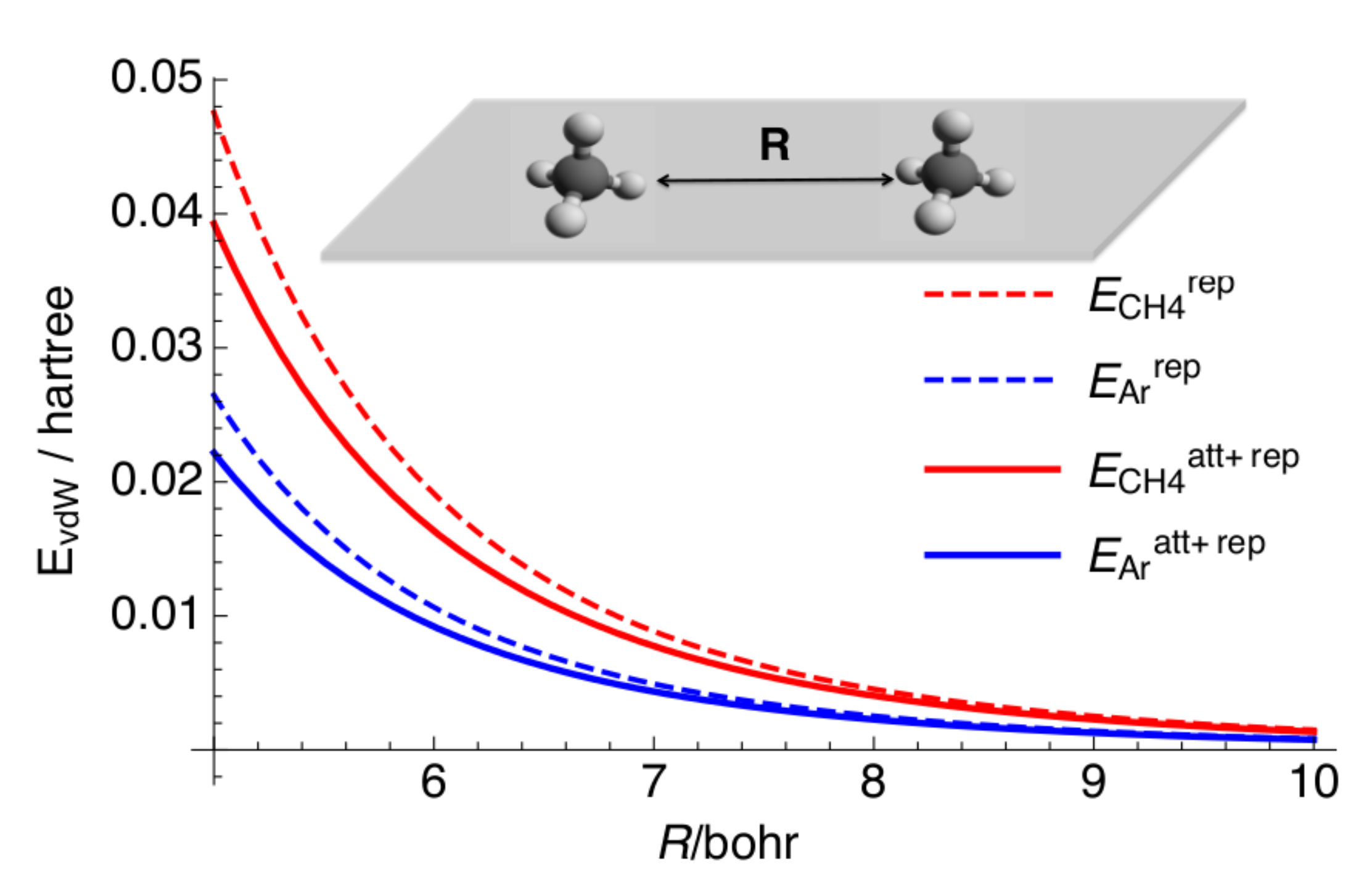}
\caption{Repulsive ground state interaction $E^{rep}$ (solid lines) and the sum of repulsion and London attraction ($E^{att}$) energy (broken lines) for argon and methane dimers on a perfectly reflecting surface.} \label{fig:methanear}
\end{center}
\end{figure}

Note that the long-range repulsive energies in quasi-1D (Eq.~\eqref{e1_4_1d}) and quasi-2D confinement (Eq.~\eqref{e1_4_2d}) are outcomes of first-order perturbation over dipolar-correlated states. {Same quantity for} a pair of asymmetric 3D-oscillators exhibit long-range repulsion that varies as $C_{5}/R^{-5}$ with
$C_5 = \frac{9}{4}\lp \frac{q^2}{m \omega^2}\rp\lp\frac{ \hbar}{m \omega} \rp \hbar \omega$, providing a generalization of the 1D and 2D confined oscillator cases. Using the parameters of methane molecule and Ar atom~\cite{Jones2013}, we obtain an estimate of long-range inter-molecular lateral potential (Fig.~\ref{fig:methanear}). As expected, the interaction is repulsive in the long range for both cases.
Note that the model presented here assumes ideal confinement. For realistic surfaces, effects of different adsorption heights, tunneling penetration of molecular electron density through the confining boundary, and the possibility of imperfect reflection need to be considered. Moreover, the interaction is purely attractive in the absence of confinement. Therefore, we expect a crossover from attraction to repulsion depending on the polarizabilty of the adsorbate molecule and the penetration depth of the adsorbent electron density, having observable effects on the surface structure of the adsorbed molecules. 

{We would now like to examine the relevance of the present analyses in realistic experimental situations where long-range repulsions or a significant decrease in attractive van der Waals interaction in presence of confinements are already reported. To connect the presented 1D analysis to experimental findings, we note that it has been found that the flow rate of water through carbon nanotubes (CNT) increases significantly with decrease in the CNT diameter \cite{Mattia2015266}. Classical molecular dynamics simulations ~\cite{Kannam2013} and DFT calculations (see Ref.~\cite{Striolo2016} for a contemporary review) underestimates these effects by several orders of magnitude. The repulsion presented in this work are missing in classical force fields, would introduce an intermolecular repulsion, which may rationalize a higher flow-rate with increasing confinement. 
The 2D oscillator dimer is related to a recent experiment~\cite{Algara-Siller2015} where a previously unknown ice-structure has been discovered when water molecules are encapsulated between two graphene sheets, thereby suggesting peculiar intermolecular interaction between water molecules under quasi-2D confinement. In fact, a recent quantum Monte Carlo study~\cite{Michaelides2016} of stable square ice between graphene sheets, shows that most of dispersion-corrected DFT functionals overestimate the binding in the water layer.
FInally, the long-range repulsion between a quasi-3D oscillator dimer presented here may provide an alternate interpretation of lateral repulsion between monolayer structures of hexabenzocoronene on Au(111), where a 90\% screening of lateral vdW interactions by the surface were suggested previously~\cite{Wagner2010}.}

In summary, we solved the problem of two oscillators coupled with a Coulomb potential under 1D, 2D, and quasi-3D confinement, utilizing a novel perturbation expansion based on Ref.\cite{Jones2013} and correlated dipolar oscillator states. The breaking of spherical symmetry yields a long-range repulsive interaction, suggesting that this is a general phenomenon for confined dimers. Our findings can be extended to systems such as atoms in optical traps, molecules confined in nanopores and molecules adsorbed on polarizable surfaces. While the presented analytic results are valid for vdW dimers, Coulomb-induced many-body effects in many-particle systems might turn out to be even more intricate, and could be addressed by extending the developed perturbation expansion on top of the microscopic many-body dispersion Hamiltonian~\cite{Distasio2014, Tkatchenko2012}.

MS thanks Jan Hermann, Dr. Igor Poltavskyi and Prof. Frederick R. Manby for many helpful and stimulating discussions.
\bibliographystyle{apsrev4-1}

\end{document}


\title{Long-Range Repulsion Between Spatially Confined van der Waals Dimers}
	\author{Mainak Sadhukhan}
	\author{Alexandre Tkatchenko}
	\email{alexandre.tkatchenko@uni.lu}
	\affiliation{Physics and Materials Science Research Unit, University of Luxembourg, L-1511 Luxembourg}
	\date{\today}
\maketitle
\section{Dipole and full-Coulomb coupling in 1D oscillator dimer}
\label{1dosc}
 The Hamiltonian describing the quantum motion of an individual Drude oscillator is 
\eq
 \ham_{0} = \frac{\p^2}{2m} + \frac{k}{2}x^2
\qe
having eigenvalues and eigenstates 
\eq
 E_{n} = \left( n + \frac{1}{2}\right) \hbar \omega 
\qe
and 
\eq
 \langle x \ket{n} = \frac{1}{\sqrt{2^n \fact{n}}}\left( \frac{m \omega}{\pi \hbar}\right)^{1/4}e^{-\left[\frac{m \omega x^2}{2 \hbar}\right]} H_{n}\left( \sqrt{\frac{m \omega}{\hbar}}x \right),
\qe  
 repectively. Here $H_{n}$ is $n$-th order Hermite polynomial. 
The total Hamiltonian for full-Coulomb coupled 1D Drude oscillator dimer is 
\eq
\ham_{tot} = \ham_{0}^{1} + \ham_{0}^{2} + V_{int} 
\qe
where the Coulomb interactions between charges are given by
\eq
V_{int} = q^2
\left[ \frac{1}{R} + \frac{1}{R + x_{2} -x_{1}} -\frac{1}{R+x_{2}} -\frac{1}{R-x_{1}} \right]
\qe
where $x_1$ and $x_2$ are coordinates of individual oscillators. For $R \gg x_1, x_2$, 
\begin{eqnarray}\label{vint_1d}
V_{int} &=& \frac{q^2}{R} \sum_{n = 1}^{\infty}\left[ \left(\frac{x_1-x_2}{R}\right)^{n} -\left( \left( \frac{-x_2}{R}\right)^n + \left( \frac{x_1}{R}\right)^n \right)\right] \nonumber \\ 
        &=&  \sum_{n = 2}^{\infty} V_{n} = \frac{q^2}{R} \sum_{n = 2}^{\infty}\sum_{k = 1}^{n-1} \binom{n}{k} \frac{x_{1}^{n-k} (-x_{2})^{k}}{R^n}
\end{eqnarray}
For $n = 2$ we obtain the dipole potential $V_{dip}$ ,
\eq
V_{2}= V_{dip} = -\frac{2q^2 x_1 x_2}{R^3} .
\qe 

The total Hamiltonian for a dipole-coupled Drude oscillator dimer is,
\begin{eqnarray}\label{dip_coupled_osc}
\ham_{tot} &=& \ham_{0}^{1} + \ham_{0}^{2} + V_{dip} \nonumber \\
 &= &\ham_{0}^{1} + \ham_{0}^{2}  + \gamma x_1 x_2  
\end{eqnarray}
where $\gamma = \frac{-2q^2}{R^3}$. $\ham_{0}^{1}$ and $\ham_{0}^{2}$ are Hamiltonians for two individual non-interacting oscillators. 

In collective coordinates 
\begin{eqnarray}
a_1 &=& \obrt \lp x_1 + x_2\rp \label{coll_coord1}\\
a_2 &=& \obrt \lp x_1 - x_2 \rp \label{coll_coord2}
\end{eqnarray}
the solution of Eq.\eqref{dip_coupled_osc} can be written as a sum of two independent oscillators with mass $m$ and frequencies $\omega_1 = \omega / f_{-}$ and $\omega_2 = \omega / f_{+}$, respectively where
\begin{equation}\label{f1f2definition}
f_{\pm} = \lp1\pm\frac{2q^2}{\mu\omega^2 R^3}\rp^{-1/2}.
\end{equation}
corresponding to $a_1$ and $a_2$, respectively. 
\section{Beyond-dipole perturbation theory in 1D}
 Now we will use $V_{int}-V_{dip}$ as perturbaton to the dipole coupled states described by Eq.\eqref{dip_coupled_osc}
Therefore, the perturbing potential is 
\eq\label{vpert}
V_{pert} = V_{int}-V_{dip} = \frac{q^2}{R} \sum_{n = 3}^{\infty}\sum_{k = 1}^{n-1} \binom{n}{k} \frac{x_{1}^{n-k} (-x_{2})^{k}}{R^n}
\qe 
\subsection{First order perturbation}
The expression for first-order perturbation correction to energy of a dipole coupled dimer is 
\eq \label{pert1}
E^{(1)} = \bra{0} V_{pert} \ket{0}
\qe
where $\ket{0}$ is ground state of dipole-coupled oscillators 
\eq\label{dip_osc_wf}
\Psi_{dip}(a_1, a_2) = \lp \frac{m \sqrt{\omega_1 \omega_2}}{\pi \hbar}\rp^{1/2} e^{-\frac{m \omega_1 a_{1}^{2}}{2 \hbar}}e^{-\frac{m \omega_2 a_{2}^{2}}{2 \hbar}}
\qe
which is the product of two gaussians in $a_1$ and $a_2$. We will analyse first two leading terms of Eq.\eqref{vpert} corresponding to $n = 3$ and $n=4$. 

\begin{itemize}
\item $n = 3$: \\
\eq
 V_{3} = \frac{q^2}{R^4}\left[ 3 x_{1} x_{2} \lp x_{2} -x_{1} \rp \right]
\qe
Using collective coordinates Eqs.\eqref{coll_coord1}-\eqref{coll_coord2} we obtain
\eq\label{v3}
V_{3} = \frac{-3 q^2}{\sqrt{2} R^4} \lp a_{1}^{2} a_{2} - a_{2}^{3} \rp
\qe
Both terms in Eq.\eqref{v3} contains odd power of $a_{2}$. As a result, the integrals in Eq.\eqref{pert1} will vanish identically owing to the fact that odd moments of a gaussian is zero. \\

\item $n = 4$:\\
We therefore consider the next term in the Eq.\eqref{vpert} corresponding to $n = 4$
\eq
 V_{4} = \frac{q^2}{R^5}\left[ 6 x_{1}^{2} x_{2}^{2} -4\lp x_{1}^{3} x_{2} + x_{2}^{3} x_{1} \rp\right]
\qe
Expressing the $x_1$ and $x_2$ in terms of $a_1$ and $a_2$, we obtain 
\eq
 V_{4} = \frac{q^2}{2R^5} \lp 7a_2^4 - a_1^4 - 6 a_1^2 a_2^2 \rp
\qe
Subsequently we obtain the first-order perturbation energy (denoted by the superscript) from potential energy contribution from $V_4$ (denoted by the subscript) of the energy using Eq.\eqref{dip_osc_wf} in Eq.\eqref{pert1} as
\eq\label{e4_1}
 E^{(1)}_{4} = \frac{3q^2}{8R^5}\lp \frac{\hbar}{m \omega}\rp^2 \left[ 7 f_{+}^2 - f_{-}^2 - 2 f_{+}f_{-} \right]
\qe
After expanding $f_{\pm}$ binomially we recover leading first-order perturbative corrections over the dipole energy as, 
\begin{eqnarray}\label{e1d}
 E^{(1)}_{4} &=&\sum_{i=1}^{\infty} \varepsilon_{i}  \nonumber \\
             &=& 2 \frac{\alpha_2 \hbar \omega}{R^5} - 8 \frac{\alpha_1 \alpha_2 \hbar \omega}{R^8} + \dots
\end{eqnarray}
containing 
the dipole polarizability 
\eq
\alpha_1= \frac{q^2}{m\omega^2}
\qe
and the quadrupole polarizability 
\eq
\alpha_{2} = \frac{3}{4}\lp \frac{q^2}{m \omega^2}\rp \lp \frac{\hbar}{m \omega}\rp 
\qe
of the Drude oscillator. The leading term of the energy is
\eq
 \varepsilon_1 = 2 \frac{\alpha_2 \hbar \omega}{R^5}
\qe
 
\end{itemize}
\subsection{Second order perturbation}
\begin{itemize}
 \item $n = 3$:\\
 The leading second order perturbation term due to $V_{3}$ is
\eq
E^{(2)}_{3} = -\left[ \frac{\vert \bra{0_{1} 0_{2}} V_{3} \ket{0_{1} 1_{2}}\vert^2}{\hbar \omega_{2}} + \frac{\vert \bra{0_{1} 0_{2}} V_{3} \ket{1_{1} 0_{2}}\vert^2}{\hbar \omega_{1}}\right]
\qe
where the $\ket{l_1 n_2}$ denotes $l$ and $n$-th excited states of oscillator $1$ and $2$, respectively. 
Using the expression for $V_{3}$ in Eq.(\ref{v3}), we find that the second term will vanish identically following the symmetry of the integrals ($a_{1}$ and $a_{2}$ have even and odd symmetry in $V_{3}$, respectively.). Therefore, the second-order perturbation correction due to $V_{3}$ is  
\eq
  E^{(2)}_{3} = -\frac{9q^4}{16 m\omega^2 R^8}\lp \frac{\hbar}{m \omega}\rp^{2}\left[  3f_{+}^2 - f_{+}f_{-}\right]^2
\qe
Expanding $f_{\pm}$ binomially, we obtain the leading order term as 
\eq
 \eta_1 = -\frac{3 \alpha_{2}\alpha_{1}\hbar \omega}{R^8}
\qe
\end{itemize}

 \section{Multipole expansion in 2 dimension}
In 2-dimension, two oscillators live in $X-Y$ plane, $X$ being the line joining origins of individual reference frames of two 2D oscillators. 

The Coulomb potential between these two 2D oscillators is
\eq\label{vcoul2d}
V_{int} =q^2\left[ \frac{1}{R} + \frac{1}{\abs{\mathbf{R}+\mathbf{r}_{2}-\mathbf{r}_{1}}} -\frac{1}{\abs{\mathbf{R}+\mathbf{r}_{2}}} -\frac{1}{\abs{\mathbf{R}-\mathbf{r}_{1}}}\right]
\qe
Our choice of coordinate system dictates 
\begin{eqnarray}
\mathbf{R} &=& (R, 0)\nonumber\\
\mathbf{r}_{1} &=& (r_{1}\cos\theta_{1},r_{1} \sin\theta_{1})\nonumber\\
\mathbf{r}_{2} &=& (r_{2}\cos\theta_{2},r_{2} \sin\theta_{2}) \nonumber\\
\mathbf{r}_{1} . \mathbf{R} &=& r_{1}R\cos{\theta_{1}}\, ; \, \mathbf{r}_{2} . \mathbf{R} = r_{2}R\cos{\theta_{2}}\nonumber
\end{eqnarray}

We will employ Laplace expansion of Coulomb potential
\eq\label{laplaceexp}
\frac{1}{\abs{{\bf r}_{1}-{\bf r}_{2}}}=\frac{1}{\sqrt{r_{1}^2+ r_{2}^2 - 2 r_{1}r_{2}\cos{\theta_{12}}}}=\sum_{l=0}^{\infty}\frac{r_{1}^{l}}{r_{2}^{l+1}}P_{l}(\cos{\theta_{12}})
\qe
where $\theta_{12}$ represents the angle between vectors $\mathbf{r_{1}}$ and $\mathbf{r_{2}}$. $P_{l}(x)$ is the Legendre polynomial of degree $l$. The formula for Legendre polynomial of degree $n$:
\eq
P_{n}(x) = 2^{n}\sum_{k = 0}^{n}\binom{n}{k} \binom{\frac{n+k-1}{2}}{n}x^{k}
\qe

Applying Eq.\eqref{laplaceexp} in Eq.\eqref{vcoul2d} we obtain

\begin{multline}
V_{int} = \sum_{l=1}^{\infty} V_{l} = \frac{q^2}{R}\sum_{l=1}^{\infty} [ \frac{\abs{\mathbf{r}_{1} - \mathbf{r}_{2}}^{l}}{R^{l}}P_{l}(\cos{\theta_{Rs}})  \\
-\frac{\lp -r_{2}\rp^{l}}{R^{l}}P_{l}(\cos{\theta_{2}})-\frac{r_{1}^{l}}{R^{l}}P_{l}(\cos{\theta_{1}}) ]
\end{multline}

 Using the expansion of Legendre polynomial,
\begin{multline}
V^{l} = \frac{2^{l}q^{2}}{R^{l+1}}\sum_{k=0}^{l}\binom{l}{k}\binom{\frac{l+k-1}{2}}{l} \\
\left[ (\Delta r)^{l-k} (\Delta x)^{k} - (-1)^{l} r_{2}^{l-k} x_{2}^{k} -r_{1}^{l-k} x_{1}^{k}\right]
\end{multline}
where $\Delta r = \abs{\mathbf{r}_{1}-\mathbf{r}_{2}}$, $\Delta x = (x_{1}-x_{2})$. \\
Legendre polynomial is an alternating odd-even parity function. As a result, if $l$ is even (or, odd), non-vanishing $k$ are also even (or, odd). 

Therefore, if $l = 2p$ , $k = 2t$, 
\eq\label{veven}
\begin{split}
V_{2p}& = \frac{4^{p}q^{2}}{R^{2p+1}}\sum_{t=0}^{p}\sum_{s=0}^{p-t}\binom{2p}{2t}\binom{\frac{2(p+t)-1}{2}}{2p}\binom{p-t}{s} \\
&\left[ (\Delta x)^{2(p-s)} (\Delta y)^{2s} - x_{2}^{2(p-s)} y_{2}^{2s} -x_{1}^{2(p-s)} y_{1}^{2s}\right]
\end{split}
\qe

and if $l = 2p + 1$ , $k = 2t +1 $,

\eq\label{vodd}
\begin{split}
V_{2p+1} &= \frac{2^{2p+1}q^{2}}{R^{2p+2}}\sum_{t=0}^{p}\sum_{s=0}^{p-t}\binom{2p+1}{2t+1}\binom{\frac{2(p+t)+1}{2}}{2p+1}\binom{p-t}{s}\\
 & \left[ (\Delta x)^{2(p-s)+1} (\Delta y)^{2s} + x_{2}^{2(p-s)+1} y_{2}^{2s} -x_{1}^{2(p-s)+1} y_{1}^{2s}\right]
\end{split}
\qe

\begin{itemize}
\item $l = 1$:\\
  \par Realizing, $P_{1} (x)= x$ we find that this term vanishes identically.
\item $l = 2:$\\
\par This is the dipole potential 
\begin{eqnarray}
V_{2} &=& \frac{q^2}{R^3}(\abs{\mathbf{r}_{1} - \mathbf{r}_{2}}^{2}P_{2}(\cos{\theta_{Rs}})
- r_{2}^2 P_{2}(\cos{\theta_{2}})-r_{1}^2P_{2}(\cos{\theta_{1}})) \nonumber \\
 &=& \frac{-2q^2}{R^3}\lp x_{1}x_{2}\rp  + \frac{2\tilde{q}^2}{R^3} \lp y_{1}y_{2} \rp
\end{eqnarray}
where $\tilde{q} = q/\sqrt{2}$.\\

\item $l = 3:$\\
 The third order potential
\eq\label{v32d_1}
V_{3} = \frac{q^2}{R^4}\lp\abs{\mathbf{r}_{1} - \mathbf{r}_{3}}^{3}P_{3}(\cos{\theta_{Rs}}) + r_{2}^3 P_{3}(\cos{\theta_{2}})-r_{1}^3 P_{3}(\cos{\theta_{1}})\rp
\qe
Let us define the collective coordinates as before,
\begin{eqnarray*}
a_{1} &=& \obrt ( x_{1} + x_{2})\\
a_{2} &=& \obrt ( x_{1} - x_{2})\\
b_{1} &=& \obrt ( y_{1} + y_{2})\\
b_{2} &=& \obrt ( y_{1} - y_{2})\\
\end{eqnarray*}
Use of these collective coordinates in Eq.\eqref{v32d_1} yields
\eq\label{v3_2d_1}
V_{3} = V_{3, x} - \frac{3q^2}{2\sqrt{2}R^{4}}\left[ 3 a_{2}b_{2}^{2} - a_{2}b_{1}^{2} - 2a_{1}b_{1}b_{2} \right]
\qe
where $V_{3, x}$ is same as $V_{3}$ in Eq.\eqref{v3}. If we define a $Y-$directional third order potential 
\eq
V_{3, y} = \frac{3\tilde{q}^2}{2R^{4}} \lp b_{2}^{3}-b_{1}^{2}b_{2}\rp
\qe

we can write the 2D, 3rd order potential as 
\begin{multline}\label{v3_2d_2}
V_{3} =V_{3, x} + V_{3, y} -  \\ 
\frac{3\tilde{q}q}{2R^{4}} \left[ 2b_{2} \lp a_{2}b_{2}-b_{1}a_{1}\rp -\lp a_{2} + b_{2}\rp \lp b_{1}^2 - b_{2}^2\rp\right]
\end{multline}
\item $l = 4:$\\
The fourth order potential in collective coordinates is
\begin{multline}\label{v4_2d}
V_{4} = V_{4, x} + \frac{3}{4}V_{4, y} - \\\frac{3q^2}{2R^{5}}\left[ 7a_{2}^2 b_{2}^{2}-4a_{1} b_{1}a_{2}b_{2} -a_{1}^2 b_{1}^{2}- a_{1}^2 b_{2}^{2}-a_{2}^2 b_{1}^{2} \right]
\end{multline}
where 
\eq
V_{4, x} = \frac{q^2}{R^{5}} \left[ 6 x_{1}^{2}x_{2}^{2}-4 \lp x_{1}^{3} x_{2} + x_{2}^{3} x_{1} \rp\right]
\qe
and
\eq
V_{4, y} = \frac{\tilde{q}^2}{R^{5}} \left[ 6y_{1}^{2}y_{2}^{2}-4\lp y_{1}^{3}y_{2} + y_{2}^{3}y_{1} \rp\right]
\qe

\end{itemize}
\section{Dipole solution in 2D}

The dipole solution of 2D oscillators are just the superposition of two independent dipole solutions differring by their charge (coupling constant).
As a result, the correlated eigenstates are defined by four independent oscillators, two in each directions having frequencies 
\begin{eqnarray}
\omega_{1}^{x} &=& \omega \lp 1- \frac{2q^2}{m\omega^{2} R^{3}} \rp^{1/2}\\
\omega_{2}^{x} &=& \omega \lp 1+ \frac{2q^2}{m\omega^{2} R^{3}} \rp^{1/2}\\
\omega_{1}^{y} &=& \omega \lp 1+ \frac{2\tilde{q}^2}{m\omega^{2} R^{3}} \rp^{1/2}\\
\omega_{2}^{y} &=& \omega \lp 1- \frac{2\tilde{q}^2}{m\omega^{2} R^{3}} \rp^{1/2}\\
\end{eqnarray}
where the superscripts indicate there direction. \\ 
We also define the directional dipole-polarizabilities 
\begin{eqnarray}
\alpha_{1}^{x} &=& \frac{q^2}{m \omega^2}\\
\alpha_{1}^{y} &=& \frac{\tilde{q}^2}{m \omega^2}
\end{eqnarray}
 The wavefunction of a 2D Drude oscillator interacting through dipole interaction is 
\eq
\ket{\psi_{\mathbf{n}}}= \ket{\phi_{n_{x}}}\ket{\phi_{n_{y}}}
\qe 
where, $ \ket{\phi}$ is one dimensional dipole solution and  
\eq
\abs{\mathbf{n}} = n_{x} + n_{y}
\qe
As before, we also define four dimensionless quantities 
\begin{eqnarray}
\frac{1}{\omega_{1}^{x}} = \frac{f_{+}^{x}}{\omega}&=& \frac{1}{\omega} \lp 1- \frac{2q^2}{m\omega^{2} R^{3}} \rp^{-1/2}\\
\frac{1}{\omega_{2}^{x}} = \frac{f_{-}^{x}}{\omega}&=& \frac{1}{\omega} \lp 1+ \frac{2q^2}{m\omega^{2} R^{3}} \rp^{-1/2}\\
\frac{1}{\omega_{1}^{y}} = \frac{f_{+}^{y}}{\omega}&=& \frac{1}{\omega} \lp 1+ \frac{2\tilde{q}^2}{m\omega^{2} R^{3}} \rp^{-1/2}\\
\frac{1}{\omega_{2}^{y}} = \frac{f_{-}^{y}}{\omega}&=& \frac{1}{\omega} \lp 1- \frac{2\tilde{q}^2}{m\omega^{2} R^{3}} \rp^{-1/2}
\end{eqnarray}
\section{Beyond dipole perturbation theory for 2D dipole-correlated Drude oscillators}
\subsection{1st order perturbation theory}
The leading order term in 1st order perturbation will come from $V_4$. However, we only need to calculate the remaining term other than the pure $x$ and $y$-component contribution since we can directly get them from our 1D results.
\eq\label{e12d4}
E^{(1)}_{2D, 4} = E^{(1)}_{x, 4}+\frac{3}{4}E^{(1)}_{y, 4} + E^{(1)}_{xy, 4}
\qe
$E^{(1)}_{xy, 4}$ in Eq.\eqref{e12d4} comes from the last term (interaction between $x$ and $y$ directional oscillators) of Eq.(\ref{v4_2d}). Since the ground state of the 2D oscillator has even parity symmetry, no term containing odd powers of any of the four variables would survive. Hence
\eq
E^{(1)}_{xy, 4} = - \frac{3q^2}{8R^{5}}\lp\frac{\hbar}{m \omega}\rp^{2}\left[  7f_{-}^{x} f_{-}^{y}- f_{+}^{x} f_{+}^{y} - f_{+}^{x} f_{-}^{y} - f_{-}^{x} f_{+}^{y}\right]
\qe
This is an attractive correction. In the leading order this term cancels the $x$-component which is repulsive. Therefore what remains is the $y$-component term
\eq
 E^{(1)}_{2D, 4} \approx \frac{3}{4}\varepsilon_{1, y} = \frac{3}{2} \frac{\alpha_{2}^{y} \hbar \omega}{R^5}
\qe 
where 
\eq
\alpha_{2}^{y} = \frac{3}{4}\lp \frac{\tilde{q}^2}{m \omega^2}\rp \lp \frac{\hbar}{m \omega}\rp 
\qe
\section{Anisotropic Half-oscillator on a surface}
The gound state of an asymmetric 3D oscillator placed on $XY$ plane which is impermeable for $z \le 0$ is
\eq
\bra {\mathbf{r}}\mathbf{0}\rangle = \phi_{0}(x) \phi_{0}(y) \psi_{0}(z)
\qe 
We will use usual symmetric oscillators along $X$ and $Y$ direction but for $Z$ direction we have to use assymmetric oscillator since the surface must be impermeable to $Z$-directional oscillator. We model the potential of $Z$-directional oscillator using a half-oscillator potential for $Z$-dimensional oscillator. 
\[ V(z) = \begin{cases} 
      \infty & \textrm{; } z \le 0  \\
      \frac{m \omega^2}{2} z^2& \textrm{; } z > 0 \\
   \end{cases} \]
The boundary condition for such oscillator will be 
\eq
\psi_{n}(0) = 0 \textrm{;  }\forall n 
\qe 
As a result, we obtain the wavefunction of the $Z$-component wavefunction 
\eq
\psi_{n}(z) = N_{n} H_{2n+1}(\sqrt{m \omega/\hbar }z)e^{-\frac{m\omega z^2}{2\hbar}} \Theta(z)
\qe 
where the normalization constant 
\eq
N_{n} = \frac{1}{\sqrt{4^{n} \fact{(2n+1)} }}\lp \frac{m \omega}{\pi\hbar}\rp^{1/4}
\qe
The ground state in $Z$-direction is 
\eq
\psi_{0}(z)= \lp \frac{m \omega}{\pi\hbar}\rp^{1/4}2 \sqrt{m \omega/\hbar} z e^{-\frac{m\omega z^2}{2\hbar}} \Theta(z)
\qe
and the ground state in $X$-direction (same as $Y$-direction) is
\eq
\phi_{0}(x)= \lp \frac{m \omega}{\pi\hbar}\rp^{1/4} e^{-\frac{m\omega x^2}{2\hbar}} 
\qe
Therefore, the expression for ground state of 3D oscillator is 
\eq
\Psi_{0}(x, y, z)= N_{0} e^{-\frac{m\omega \lp x^2 + y^2 +z^2 \rp}{2\hbar}}z  \Theta(z)
\qe
where the normalization constant is 
\eq
N_{0} = 2 \sqrt{m \omega/\hbar} \lp \frac{m \omega}{\pi\hbar}\rp^{3/4}
\qe
The total energy of a pair of Drude oscillator is given by the same expression as in Eq(\ref{vcoul2d}) only now the vectors are 3D and $\mathbf{R} = (R, 0, 0)$.

Therefore the total induction interaction energy (first order perturbation correction) is given by 
\eq
E_{int}= q^{2}\left[ \frac{1}{R} + E_{nn} - E_{np} -E_{pn}\right]
\qe
where, 
\eq
E_{nn} = \bra{\mathbf{0}_{1}\mathbf{0}_{2}}\frac{1}{\abs{\mathbf{R}+\mathbf{r}_{2}-\mathbf{r}_{1}}}\ket{\mathbf{0}_{1}\mathbf{0}_{2}}
\qe
\eq
E_{np} = \bra{\mathbf{0}_{2}}\frac{1}{\abs{\mathbf{R}+\mathbf{r}_{2}}}\ket{\mathbf{0}_{2}}
\qe
and
\eq
E_{pn} = \bra{\mathbf{0}_{1}}\frac{1}{\abs{\mathbf{R}-\mathbf{r}_{1}}}\ket{\mathbf{0}_{1}}
\qe
Expressing the Coulomb kernel in terms of Gaussian integral using the identity,
\eq\label{heatkernel}
\frac{1}{r} = \frac{2}{\sqrt{\pi}}\int_{0}^{\infty} e^{-s^{2}r^{2}} \diff s
\qe
we obtain the expression for $E_{nn}$ ($\beta = \sqrt{m \omega/\hbar}$)
\eq
E_{nn} =  \frac{2}{\sqrt{\pi}} N_{0}^{4} \int_{0}^{\infty}\diff s  I_{x} I_{y} I_{z}
\qe
The $X$ integral
\begin{eqnarray}
I_{x} &=& \int_{-\infty}^{\infty} \diff x_{1}\int_{-\infty}^{\infty} \diff x_{2} e^{-\left[ \beta^{2} (x_{1}^{2}+x_{2}^{2}) + s^{2}(R +x_{2}-x_{1})^{2}\right]}\nonumber \\
 &=& \frac{\pi e^{-\frac{R^{2}\beta^{2}s^{2}}{2s^{2} + \beta^{2}}}}{\beta \sqrt{2s^{2}+ \beta^{2}}}
\end{eqnarray}
The $I_{y}$ can be readily obtained by setting $R = 0$.
\eq
I_{y} = \int_{-\infty}^{\infty} \diff y_{1}\int_{-\infty}^{\infty} \diff y_{2} e^{-\left[ \beta^{2} (y_{1}^{2}+y_{2}^{2}) + s^{2}(y_{2}-y_{1})^{2}\right]}=\frac{\pi}{\beta \sqrt{2s^{2}+ \beta^{2}}}
\qe
The $Z$-integral yields
\begin{eqnarray}
I_{z} &=& \int_{-\infty}^{\infty} \diff z_{1}\int_{-\infty}^{\infty} \diff z_{2} e^{-\left[ \beta^{2} (z_{1}^{2}+z_{2}^{2}) + s^{2}(z_{2}-z_{1})^{2}\right]}(z_{1}z_{2})^{2}\Theta^{2}(z_{1})\Theta^{2}(z_{2})\nonumber\\
&=& \frac{\pi}{16 \beta^{5}} \frac{\lp 3s^{4}+2s^{2}\beta^{2} +\beta^{4} \rp}{\lp 2s^{2}+ \beta^{2}\rp^{5/2}}
\end{eqnarray}
Hence 
\eq\label{enn}
E_{nn} =  \frac{2\beta^{3}}{\sqrt{\pi}}  \int_{0}^{\infty}\diff s  \frac{\lp 3s^{4}+2s^{2}\beta^{2} +\beta^{4} \rp}{\lp 2s^{2}+ \beta^{2}\rp^{7/2}} e^{-\frac{R^{2}\beta^{2}s^{2}}{2s^{2} + \beta^{2}}}
\qe
Similarly we obtain
\eq\label{epn}
E_{pn} = \frac{2\beta^{5}}{\sqrt{\pi}}\int_{0}^{\infty}\diff s  \frac{e^{-\frac{R^{2}\beta^{2}s^{2}}{s^{2} + \beta^{2}}}}{\lp s^{2}+ \beta^{2}\rp^{5/2}} 
\qe
and 
\eq
E_{pn} = E_{np}
\qe
The numerical integration of Eq.\eqref{enn} and Eq.\eqref{epn} followed by dimensional analysis for $\{q, m, \omega, R\}$ provides the leading order interaction as $C_{5}/R^{-5}$ with
\eq
C_5 = \frac{9}{4}\lp \frac{q^2}{m \omega^2}\rp\lp\frac{ \hbar}{m \omega} \rp \hbar \omega
\qe